\documentclass[12pt,preprint]{aastex}
\usepackage{graphicx}
\usepackage{color}

\begin{document}
\title{Chandra Survey of Radio-quiet, High-redshift Quasars} 
\author{Jill Bechtold\altaffilmark{1}, Aneta Siemiginowska\altaffilmark{2},
Joseph Shields\altaffilmark{3}, 
Bozena Czerny\altaffilmark{4}, Agnieszka Janiuk\altaffilmark{4}
Fred Hamann\altaffilmark{5},
Thomas L. Aldcroft\altaffilmark{2}, Martin Elvis\altaffilmark{2}, 
Adam Dobrzycki\altaffilmark{2}
}

\altaffiltext{1}{Steward Observatory, University of Arizona, 933 N. Cherry 
Avenue, Tucson AZ 85721;
jbechtold@as.arizona.edu}
\altaffiltext{2}{Harvard-Smithsonian Center for Astrophysics; asiemiginowska,aldcroft,elvis,adam@cfa.harvard.edu}
\altaffiltext{3}{Department of Physics and Astronomy, Ohio University, Athens, OH 45701; shields@phy.ohiou.edu}
\altaffiltext{4}{Nicolaus Copernicus Astronomical Center, 
Warsaw, Poland; agnes@camk.edu.pl,bcz@camk.edu.pl}
\altaffiltext{5}{Department of Astronomy, University of Florida, 211 Brayant Space Science Center, Gainesville, FL 32611, hamann@astro.ufl.edu}

\received{}
\accepted{}

\begin{abstract}
We observed 17 optically-selected, radio-quiet high-redshift 
quasars with the Chandra Observatory ACIS, and detected 16 of them.  
The quasars have redshift between 3.70 and 6.28 and include the highest
redshift quasars known.
When compared to low-redshift quasars 
observed with ROSAT, these high redshift quasars are 
significantly more X-ray quiet.
We also find that the X-ray spectral index of the high redshift objects is
flatter than the average at lower redshift.  These 
trends confirm 
the predictions of 
models where the accretion flow is described by
a cold, optically-thick accretion disk surrounded by a hot, optically thin 
corona, provided the viscosity parameter $\alpha \ge$ 0.02. The high redshift
quasars have supermassive black holes with masses $\sim10^{10}$ M$_{\odot}$,
and are accreting material at $\sim$0.1 the Eddington limit. 
We detect 10 X-ray photons from the $z=6.28$ quasar SDS 1030+0524, 
which may have a Gunn-Peterson trough and be near the redshift of
reionization of the intergalactic medium.  The X-ray data  
place an upper limit on the optical depth of the intergalactic
medium $\tau(IGM) < 10^6$, compared to the lower limit from the spectrum of
Ly$\alpha$ and Ly$\beta$, which implies $\tau(IGM) > 20$. 

\end{abstract}


\section{Introduction}

Rapid growth in the number of quasars discovered with $z > 4$ has taken place
in the last few years (Warren et al. 1987; Schneider, Schmidt \& Gunn 1989; 
Kennefick et al. 1995ab; Hook, McMahon, Irwin, \& Hazard 1996; 
Storrie-Lombardi, McMahon, Irwin \& Hazard 1996; Hook \& McMahon 1998;  
Fan et al. 2000, 2001ab; Becker et al. 2001; 
Schneider et al. 2002).   We are thus now able
to probe the quasar phenomenon in a new regime of high
luminosity and youth that is important for understanding the structure
of quasars and their evolution.  At the present time, there are
about 130 QSOs identified at $z > 4$ (NED 2002), but X-ray data for these sources
remain sparse, especially for the radio-quiet majority (Bechtold et al. 1994; 
Pickering, Impey \& Foltz 1994; Henry et al. 1994; Mathur \& Elvis 1995; Siebert et al. 1996; 
Fabian et al. 1997; 
Zickgraf et al. 1997; Siebert \& Brinkman 1998; 
Schneider et al. 1998; 
Wu, Bade \& Beckmann 1999; Kaspi et al. 2000; Brandt et al. 2001; 
Vignali et al. 2001).  The X-ray continuum 
in active nuclei is thought to arise in a Comptonized wind or corona  
associated with the inner accretion disk around a super-massive black
hole (Haardt \& Maraschi 1993; Czerny \& Elvis 1987).  
Additional components of the X-ray emission are also
commonly seen in high signal to noise spectra at low redshift,
including X-rays associated with beamed synchrotron plasma;
Compton reflected emission; line emission in Fe K$\alpha$; weak 
absorption from warm ionized material; and strong absorption from
cold material (see George et al. 2000, Reeves \& Turner 2000 and references therein).  
Whether or not there is evolution in the structure of quasars 
with redshift can in principle be learned from X-ray spectroscopy of high redshift quasars.  

We report here a survey with the Chandra X-ray Observatory, of 
optically selected, radio-quiet high redshift quasars.  This 
project is part of our long-term program to characterize
the multi-wavelength emission from quasars as a function of redshift 
(Bechtold et al. 1994 ab, Kuhn et al. 2001).
The statistical properties of the ensemble population of
quasars evolve strongly with redshift:  the break point, L$^*$, of
the luminosity function shifts to higher luminosities by a factor of
$\sim$50 between z=0 and z=2 (Boyle, Fong, Shanks \& Peterson 1987).  We hope to
understand this evolution in terms of a physical model for the evolution of
individual objects:  does this evolution reflect slow changes in a few rare
objects, or a short-lived phase that most galaxies go through?
Are there differences in the spectral evolution of radio-loud
and radio-quiet quasars? Can these be used to understand the origin
of the radio emission?
The X-ray photons, whose production is closely linked to processes
in the central engine, should provide direct
clues to the physical conditions in the central few parsecs,
where most of the quasar energy is produced, and will therefore
help answer these questions.  With the recent suggestion that
reionization of the intergalactic medium 
happened at $z\sim6.3$ (Becker et al. 2001), it is possible that
$z\sim 6$ is the epoch when the first quasars were born.
Thus, the quasars targeted in our survey are among
the first objects formed in the Universe.

High redshift quasars are rare, and require pointed observations at known
objects for study; our expectation is that XMM and Chandra serendipitous
surveys will not cover sufficient area on the sky to find many of them.
The increase in sensitivity provided by Chandra and XMM however make it
possible to increase significantly the number of $z > 4$ quasars with
X-ray data with only a relatively modest investment of 
telescope time per object.

We chose targets from samples of high redshift quasars, based on
their optical luminosity.  The quasars were found in optical surveys, including 
the APM multicolor survey (Williger et al. 1994; Storrie-Lombardi et al. 
1994, 1996), the Second Palomar Observatory Sky Survey (Smith et al. 1994; 
Kennefick et al. 1995ab),  
and the Sloan Digital Sky Survey (Fan et al. 1999, 2000; Becker et al. 2001).
Subsequent observations with FIRST and NVSS confirm that are all but one are radio quiet
(Stern et al. 2000; Fan et al. 2001b; NED 2002).
Since these quasars are unusually bright, they have been the subject
of other studies, including high quality optical spectrosopy to
study the emission lines (Constantin et al. 2001) and absorption lines
(Storrie-Lombardi et al. 2001; Peroux, Storrie-Lombardi \& McMahon 2001).

We calculated observing times required to detect the quasars in X-rays
with 100 photons, if the quasar had the average optical to X-ray flux
ratio seen at lower redshift, so that
we would be assured of a detection (9-10 photons) if the quasar were
more X-ray weak.  In fact, we detected all but one quasar in the sample.
Observations of three quasars were taken from the public archive, to extend our 
sample to $z=6.28$.

\section{Observations and Analysis}

Seventeen quasars 
were observed with the Chandra X-ray Observatory and 
Advanced CCD Imaging Spectrometer (ACIS-S, Garmire et al., in
preparation; Weisskopf \& O'Dell 1997).  
All the observations were taken with the quasar on the ACIS-S3 CCD, 
and reduced with the standard pipe-line reduction software, and 
CIAO (2002; version 2.2).
Initially, we pointed off-axis to mitigate pile-up in anticipation
of high count rates, but later moved the quasar position on axis.
Pile-up is negligible for all observations in our survey.

The observations are summarized in Table 1.  We list the
exposure times, observation times and net counts detected.  For every object 
except PSS 1435+3057, we detected a significant X-ray source within
1 arcsec of the optical position, so there is no doubt to the identification.

For PSS 1435+3057, we give a 3$\sigma$ upper limit to the X-ray photons at the 
optical position.
There is a weak source (11 photons) located 3.8 arcsec from the optical position.
This source is beyond the range of aspect errors (Aldcroft 2002), but may be 
associated with an extended structure from the quasar (c.f. Schwartz 2002).

In Table 2, we list X-ray flux rates and other X-ray parameters. 
Although the number of photons detected is small,
we were able to derive meaningful spectral fits.  We used Sherpa (CIAO 2002, 
version 2.2) to
fit power law parameters to the counts, including absorption by
the Milky Way column
density of neutral hydrogen fixed at the value inferred from 21-cm
maps (COLDEN 2002, which is 
based on Dickey \& Lockman 1990).  We fit a function of the standard form

\begin{equation}
A(E) = { f_{0} [{ {E} \over {1 keV} } ] ^{-\Gamma_x}}
\end{equation}

\noindent
where $\Gamma_x$ is the energy index and f$_o$ is the normalization
at 1 keV, with units of photons keV$^{-1}$ cm$^{-2}$ s$^{-1}$. 
We restricted our fits to the range 0.3-6.5 keV, where the
calibration is reliable at the present time, and the Chandra background 
is the lowest.  The source extraction region for each quasar 
was a circle with 10 pixels (4.92 arcsec) radius. There were no background flares during the observations.   

The choice of cosmology is 
important when converting fluxes to luminosities.  
The absolute B magnitude, $M_B$, is related to apparent 
magnitude, $m_B$, by

\begin{equation}
M_B = m_B + 5 - 5~ log~ D
\end{equation}
\noindent
where

\begin{equation}
D = { {c} \over {H_{o}} } A
\end{equation}

\noindent
where c is the speed of light and H$_o$ is the Hubble constant.

In general (Peacock 1999, his equation 3.39), 
for a flat Universe, 

\begin{equation}
A = (1+z) \int_{0}^{z} ( \Omega_{\Lambda} + \Omega_M ( 1 + z )^3 ) ^{-1/2}~ dz
\end{equation}

\noindent For $q_{0} > 0$ and $\Omega_{\Lambda}$=0, we have

\begin{equation}
A = z ( 1 + { {z ( 1 - q_{0} )} \over { (1 + 2 q_{0} z)^{0.5} + (1 + q_{0} z) } })  
\end{equation}

\noindent
which simplifies to 

\begin{equation}
 A = z ( 1 + {{z} \over {2}} )  
\end{equation}

\noindent
for $q_{0} = 0$ (see also Carroll, Press \& Turner 1992).

\noindent
The quasar luminosity function is usually reported assuming 
$q_0= 1/2$,  H$_o$= 50 $km$ $s^{-1}$ $Mpc^{-1}$, $\Lambda$=0,
while observers often adopt $q_0 = 0$ and H$_o$ = 70 $km$ $s^{-1}$ $Mpc^{-1}$ 
and $\Lambda$=0.  
The $``$best" fit to cosmic microwave
background data, Type Ia SNe light curves, and large scale structure models
suggest that
$\Omega_{\Lambda}$ = 0.7, $\Omega_{M}$=0.3, and H$_o$= 70 km s$^{-1}$ Mpc$^{-1}$.
We refer to this choice of cosmological parameters as $`` \Lambda$CDM" in calculations
below.  Note that the integral of equation (4) must be evaluated numerically.
Below we report luminosities for each of these three cosmologies.

\section{The Evolution of $\bf\alpha_{\bf ox}$ for Radio-Quiet Quasars}
 
Following Zamorani et al. (1981) we compute a ratio of X-ray to 
optical flux,  $\alpha_{ox}$, where

\begin{equation}
 \alpha_{ox} = - log (f_x / f_{opt}) / log (\nu_{x} / \nu_{opt})
\end{equation}
\noindent
and log $\nu_{x}$ = 17.6845 for rest-frame 2 keV and log $\nu_{opt}$ = 15.0791 
for rest frame 2500 \AA.  

In Table 2 we give the results.  For all objects,
the continuum flux at 1450 \AA~ is available from the literature (see
Table 1), 
usually in the form of apparent AB magnitude at 1450 \AA ~ in the rest frame of
the quasar, $m_{AB}$, where

\begin{equation}
m_{AB} = -2.5 ~ log f_{\nu} - 48.57
\end{equation}

\noindent
so that $f_{\nu}$ has units of $ergs$ $s^{-1}$ $cm^{-2}$  $Hz^{-1}$.  
We compute $f_{\nu}$ at 2500 \AA, which has units of 
$ergs$ $s^{-1}$ $cm^{-2}$ $Hz^{-1}$, 
assuming the optical continuum has a power law spectral energy 
distribution with frequency spectral index $\alpha$ = -0.3.  
We compute $f_x$ at 2 keV 
in the rest frame of the quasar from the measured flux at 1 keV in
our observed frame, assuming a power law with energy index $\Gamma_x$=2.2.

Figure 1 shows $\alpha_{ox}$ versus redshift, $z$, and Figure 2 shows 
$\alpha_{ox}$ versus absolute B magnitude, M$_{B}$.   
For clarity, we plot the errors on $\alpha_{ox}$ only for the quasars with $z>5$.  
For the rest, although the number
of X-ray photons detected is small and the uncertainties on $f_x$ 
correspondingly large, the division by $log (\nu_{x} / \nu_{opt})$ = 2.60 
makes the errors only $\sim$ 0.2.

For a low redshift comparison sample, we use the compilation of 
ROSAT all-sky survey and pointed observations for radio-quiet quasars 
(Yuan et al. 1998) 
and radio-loud quasars and blazars 
(Brinkman et al. 1997), supplemented 
with observations of other high redshift quasars
from the literature (Schneider et al. 1998; 
Kaspi, Brandt \& Schneider 2000; Brandt et al. 2001; Vignali et al. 2001).  
Since Yuan et al. (1998) and Brinkman et al. (1997) 
did not list $\alpha_{ox}$, we computed it from
the listed 
unbsorbed X-ray flux density (0.1-2.4keV), assuming that the spectrum
is a power law, with energy index $\Gamma_x$ with the best fit $\Gamma_x$ listed
by Yuan et al. (1998).  
For objects with no $\Gamma$ measured, we assume $\Gamma_x$ = 2.58 for
$z<0.5$, $\Gamma_x$ = 2.46 for $0.5<z<1.0$, $\Gamma_x$ = 2.35 for $1.0<z<2.0$
and $\Gamma_x$ = 2.2 for $z>2.0$ (Yuan et al. 1998).  
For the optical flux, we used the absolute B-magnitude for each object listed in 
Veron-Cetty \& Veron (2000) which includes a K-correction.  We extrapolated to
2500 \AA~ in the rest frame of the quasar assuming a power 
law with $\alpha$ = -0.3.
The data points plotted in Figure 1 and Figure 2 are available in the
electronic version of this paper, and on our website.

The $z>4$ radio-quiet quasars are clearly more X-ray quiet than their low-redshift
counterparts, even when their extreme luminosity is taken into account.
Previous studies (e.g. Avni, Worrall \& Morgan 1995; 
Brandt et al. 2001; and references
therein) found that $\alpha_{ox}$ depends mostly on optical luminosity,
although the correlation of luminosity and redshift in the 
observed samples made it difficult to sort out whether $\alpha_{ox}$
depended mostly on optical luminosity or redshift (e.g. Bechtold et al. 1994; 
Pickering, Impey \& Foltz 1994).
The Chandra sample also suffers from a strong correlation between redshift and luminosity. 
In Figure 1 the three Chandra
quasars at $z\sim3$ are strikingly offset from the ROSAT quasars
to larger $\alpha_{ox}$, while in figure 2 they show no offset.
This is likely due to the newer surveys from
which they are taken which cover larger solid angle than those
used to select the ROSAT high z quasars.
As a result they find systematically more luminous quasars.

To quantify the result, we computed the generalized Kendall's tau,
including lower limits on $\alpha_{ox}$ for quasars not detected at
high redshift (Avni 1976; Avni, Soltan, Tananbaum \& Zamorani 1980; 
Feigelson \& Nelson 1985; 
Avni \& Tananbaum 1986;  
Isobe, Feigelson \& Nelson 1986; Akritas \& Siebert 1996).  
We used the IRAF program {\sc bhkmethod}  
to search for correlations between (1) redshift and $\alpha_{ox}$ and
(2) absolute B magnitude and $\alpha_{ox}$.  

We find that
$\alpha_{ox}$ is anticorrelated with redshift, with Kendall's tau = -0.38, 
and Z-value =8.4, so that a correlation is
present at 8.4$\sigma$ significance.  
On the other hand, the probability is 0.05 that
$\alpha_{ox}$ is not correlated with absolute B-magnitude (Z-value=1.9).
Thus, we find that $\alpha_{ox}$ depends primarily on
redshift, and weakly on luminosity.

Does this conclusion depend upon our assumptions for the X-ray spectral index,
$\Gamma _x$?  In figure 4, we plot $\Gamma _x$ versus redshift and versus
luminosity, for the Yuan et al. (1998)
ROSAT sample and our high redshift sample.  The high redshift quasars
if anything are flatter (smaller $\Gamma_x$) than their low redshift 
counterparts.  If we had adopted a mean $\Gamma _x$ of 1.5 instead of 2.2,
we would have {\it increased} $\alpha_{ox}$ by $\sim$0.3, making the 
difference between low- and high- redshift greater.  Thus the assumed
spectral indices cannot account for the result.  

Figures 1 and 2 confirm the result seen previously that 
radio-loud quasars are more
X-ray loud than radio quiet quasars, at all redshifts and luminosities.  

Although we have emphasized how X-ray quiet the high redshift quasars
are given their optical luminosity, they are still prodigious emitters
of X-ray photons.  In Table 3, we list the X-ray luminosity (erg s$^{-1}$),
in the 2 - 10 keV band, computed from the fits in Table 2.  The quasars
have X-ray luminosities of $10^{45} - 10^{47}$ erg s$^{-1}$.

\section{Comparison with Other Studies}

After submission of this paper, three papers were submitted to the {\it Astrophysical
Journal Letters} about the Chandra observations of SDS 0836+0054, SDS 1306+0356
and SDS 1030+0524. These quasars were observed with director's discretionary time and
were placed in the public archive immediately.  Brandt et al. (2002) and
Mathur, Wilkes \& Ghosh (2002) report fluxes for the three quasars and conclude, 
contrary to the results reported here, 
that the optical-to-X-ray flux ratios of these three highest redshift quasars 
are not significantly different from
those of low redshift quasars. The different conclusion is a result of the different 
comparison samples used, which in both papers were smaller than the one
presented here.  Schwartz (2002) reported not only the core fluxes for the
three quasars, but also argued that an X-ray source 23$\arcsec$ from SDS 1306+0356
with no optical counterpart on the Palomar Sky Survey is in fact associated with
the quasar. He postulates that the X-rays are
Compton scattered cosmic microwave background 
photons from a jet structure, similar to
those seen at lower redshift (Tavecchio et al. 2000; 
Celotti, Ghisellini \& Chiaberge 2001; Siemiginowska et al. 2002).  
Future observations can confirm this intriguing result.
In our sample, as discussed above, we see a similar source 3.8$\arcsec$ from
the optical position of PSS 1435+3057.

For the core X-ray fluxes, we measure slightly different values than those reported
in the three other papers.
We measure 24, 10 and 19 photons for SDS 0836+0054, SDS 1306+0356
and SDS 1030+0524, respectively. Schwartz (2002) measures
21, 6, 16 photons; Mathur, Wilkes \& Ghosh (2002) measure 21, 6, 18 photons; and
Brandt et al. (2002) measure 21, 6 and 17 photons.  The difference lies in
two factors:  the energy range used (we use 0.3 - 6.5 keV, whereas the others use
0.5 - 7 keV or 0.5 - 8 keV) and the difference in
circle radius used for source extraction (we use 4.9 $\arcsec$, whereas 
they use 1.2 - 2.9 $\arcsec$).  We chose the 0.3 - 6.5 keV range because it has the lowest
background, and the larger circle extraction to include the point spread function
at all energies.  The result is a greater number of net photons for all sources, and
a more reliable measure of the source properties.

\section{Mean X-ray Spectrum}

Figure 3 suggests that the X-ray spectral index flattens at high redshifts. 
This could be caused by two effects. 
First, as the redshift increases, the observed X-ray band samples greater and greater 
energies.  Thus, the flattening may be caused by intrinsic flattening
of the power law at high energies.  Second, the quasars in
our survey may have flatter intrinsic spectra at all energies, due to their high
luminosities, or high redshift.  

We applied Kendall's generalized tau to investigate the dependance of $\Gamma_x$ on
redshift and optical luminosity. We did not include objects for which a 
power law index had not been measured.  We found that $\Gamma_x$ is 
anti-correlated with redshift (Kendall's tau=-0.4319, Z-value = 8.46) and positively 
correlated with luminosity (Kendall's tau=0.2330, Z-value = 4.57).  The data 
suggests that $\Gamma_x$ depends strongly on both redshift and luminosity.

We note that at the redshifts of the quasars in our sample, 
the usable Chandra energy range, 0.3-6.5 keV, corresponds
to relatively hard X-rays, 1.5 keV to 32 keV.
Therefore, we expect absorption by intervening damped Ly$\alpha$ systems
and other intervening absorbers to be neglible, unless the absorbers are
at low redshift.  Since we chose the sample
objects to avoid known broad-absorption line quasars (see Green et al. 2001), 
intrinsic absorption is also probably negligible.
We discuss the special case of SDS 1030+0524 at $z=6.28$ below.

\section{Implications for M$_{\bf BH}$ and $\dot{\bf m}$}

A successful model for explaining the optical/ultraviolet 
continuum of quasars (the Big Blue Bump) involves accretion onto
a supermassive black hole through an optically thick, physically
thin disk (Shields 1978; Malkan 1983; 
Bechtold et al. 1987; Czerny \& Elvis 1987; 
Sun \& Malkan 1989).   The emitted spectrum is then the integral of Planck 
spectra of different temperatures, resulting in a flat continuum, 
$\propto \nu^{-1/3}$  through the optical-UV. There is a near exponential
falloff blueward of a cutoff energy, E$_{co}$, since there is a maximum 
temperature for the accreting material, nearest 
the black hole.  For 
the supermassive black holes in quasars accreting near the Eddington limit,
the cutoff is E$_{co}$ $\sim$ 10-100 eV.  

The spectral shape and luminosity of the accretion disk can be predicted 
and depends on five parameters: the black hole mass (M$_{BH}$);
the mass accretion rate $\dot{m}$; the total angular momentum; the
viscosity in the disk, $\alpha$; and the inclination at which we 
observe the disk.  If we make the usual assumptions 
that the black hole is maximally spinning (Kerr black hole; see 
Elvis, Risalti \& Zamorani 2002), that we 
are observing the disk face-on, and that the 
viscosity $\alpha$=0.1, then the spectral energy distribution and luminosity
depends on two parameters, M$_{BH}$ and $\dot{m}$.    

Malkan (1990) showed that for a Kerr black hole, 
an analytical expression for the emitted spectrum is 

\begin{equation}
L_{\nu} \sim A( {{\nu} \over {\nu_{co}} } )^{1/2} exp( - { {\nu} \over {\nu_{co}} } )
\end{equation}

\noindent
where A and $\nu_{co}$ are the normalization and cutoff frequency.
The cut-off energy can then be written (Wandel 2000) as

\begin{equation}
h\nu_{co} = (6 eV) \dot{m}^{1/4}M_8^{-1/4} = (20eV)L_{46}^{1/4}M_8^{-1/2}
\end{equation}

\noindent
where $M_8$ is the black hole mass, in units of $10^8$ solar masses.

Observationally, is it also possible to relate M$_{BH}$ to the width of
the broad emission lines, assuming the BLR gas is in Keplerian motion,
and calibrating the size (and hence ionization parameter) of the gas by
reverberation mapping of low-redshift AGN (Peterson \& Wandel 1999; 
Wandel, Peterson \& Malkan 1999).

Thus, with two observations --  the width of H$\beta$ and the continuum 
luminosity -- we
can solve for the two free parameters of the model, $M_{BH}$ and $\dot{m}$ 
(Wandel \& Petrosian 1988; Wandel \& Boller 1998; Wandel 1999).
In Figures 4 and 5 we plot the result.  For comparison, we also plot the 
results for 
narrow line Seyfert 1's (Crenshaw 1986; Stirpe 1990; Boller, Brandt 
\& Fink 1996; Brandt, Mathur \& Elvis 1997), 
the PG quasars (Boroson \& Green 1992; Miller et al. 1992),
and the LBQS quasars 
(Forster et al. 2001).  For high redshift quasars, H$\beta$ is in the near-IR,
and only the most luminous quasars have measurements.  We use H$\beta$ 
whenever possible (Hill, Thompson \& Elston 1993; Rokaki, Boisson, \& 
Collin-Souffrin 1992;  Nishihara et al. 1997; McIntosh et al. 1999).  
For the rest, we use the FWHM of the CIV emission line, 
and assume the relation given by Corbin (1991), 

\begin{equation}
FWHM(H\beta) = 1.35 ~ FWHM (CIV) - 1391 ~ km ~ s^{-1}
\end{equation}

\noindent
We use C IV for intermediate redshift quasars from 
the LBQS (Forster et al. 2001), 
and observations of $z\sim4$ quasars from Constantin et al. (2002).
To estimate the luminosity, we 
used the optical luminosities computed from the 
magnitudes listed in Table 1, or from the Veron-Cetty \& Veron (2000) catalog.
For all quasars, we apply a bolometric correction of 10.

For the $\Lambda$CDM cosmology (Figure 5) we see that the NLSy1's have relatively
low mass black-holes, PG quasars are more massive, and 
the high redshift quasars have very massive black holes.  The typical 
black hole mass for the $z=4-6$ quasars is 
$\sim$ 10$^{10}$ M$_{\odot}$, with most objects accreting between 
10 $^{-2}$ and 0.8 of the Eddington mass accretion rate.     

\section{Accretion disk and X-ray emitting corona}

We now use the X-ray measurements to further investigate the nature of
the black hole and accretion in quasars.  
Janiuk \& Czerny (2000) have presented calculations of the X-ray spectrum
expected from a hot corona associated with an accretion disk. The model
predicts the fraction of energy dissipated in the corona as a function of the
disk radius. The coronal dissipation is assumed to be proportional to the gas  
pressure, and the pressure at the base of the corona is determined by the
condition for the disk/corona transition. This requirement is consistent
with the evaporation/condensation equilibrium condition (Rozanska \&
Czerny 2000) and effectively means that 
bremsstrahlung and Compton cooling at the base of the corona 
are comparable. The spectrum from the disk and
the corona are computed locally, taking into account the Comptonization in the hot
coronal plasma. 

The observed model spectrum is integrated over the disk surface.  The spectrum 
is parameterized by the black hole
 mass, accretion rate and coronal viscosity, $\alpha$. Here we used a code
which includes tabulated amplification factors for Comptonization in
the corona derived from Monte Carlo simulations 
(Janiuk, Czerny \& Zycki 2000), instead of the analytical
approach used by Janiuk \& Czerny (2000).
In contrast to the calculations described 
in the previous section, we assume 
a Schwarzschild, not Kerr black hole. We calculate model spectra 
assuming the disk is face-on.  
Based on low luminosity accretion models (ADAFs,
Kurpiewski \& Jaroszynski 2000, 1999) we expect that the X-ray spectra will be
harder for the maximally rotating black hole. Kerr geometry should be considered
in the future models to quatify the effects.

We defer a detailed comparison of the new data with the models to 
a future paper, and here present only the general trends.  
In Figure 6 we plot representative spectral energy distributions in
order to illustrate the dependence on parameters.  We plot  
models with the black hole mass fixed at $10^{10}$ $M_{\odot}$
and different viscosities and accretion rates.  We see that the
X-ray spectral index is only a weak function of the parameters, but
that the optical/UV luminosity and $\alpha_{ox}$ are strong functions
of the viscosity, $\alpha$, and accretion rate.   We use the standard 
$\alpha$ viscosity prescription (Shakura \& Sunyaev 1973) which describes
the efficiency of angular momentum transfer in the disk. 

Figure 7 shows how $\alpha_{ox}$ and $\Gamma_x$ depend on 
parameters.  We plot the results for models with black hole mass 
$10^7$ and $10^{10}$ and accretion rates of 0.01 and 0.8 Eddington,
which bracket the values derived for the luminous 
quasars in Figures 4 and 5.  

\noindent
In Figure 7 we see the following trends:

\noindent
1.  The large observed scatter in $\alpha_{ox}$ 
is predicted naturally by the models, for a plausible range of
accretion rate. 

\noindent
2.  If the luminous high redshift quasars have on average more massive  
black holes than lower redshift quasars, then they are predicted to have
larger $\alpha_{ox}$. That is, the quasars in our sample are predicted to  
be more X-ray quiet than the PG or LBQS quasars, in agreement with
the observations.

\noindent
3.  For the massive black holes in the quasars in our sample, 
$\alpha_{ox}$ $\sim$ 2 implies that the viscosity parameter 
$\alpha > 0.02$. 

\noindent
4.  The predicted X-ray spectral index, $\Gamma_x$, is between 1.2 and 1.6, similar to 
what we measure for the high redshift sample.  The low redshift
quasars with steeper $\Gamma_x$ must have an extra soft component,
or more complex spectra than the simple power law fits.

The models predict the spectrum throughout the ionizing ultraviolet
and soft X-rays, and therefore can be used to predict emission line
properties, particularly the flux and equivalent width of C IV.
The implications for the broad line region will be discussed further in a 
future paper.

\section{Detection of SDS 1030+0524 at $\bf z=6.28$}

SDS 1030+0524 is the highest redshift quasar discovered to date.
Becker et al. (2001) present the ultraviolet spectrum which
shows very strong absorption just blueward of the Ly$\alpha$ 
emission line.  They convincingly argue that the absorption is 
far stronger than what is predicted from 
a simple extrapolation of the Ly$\alpha$ forest from lower redshift, 
so that they have detected 
a Gunn-Peterson trough.  They conclude that the intergalactic 
medium (IGM) was in the
process of reionization 
at $z\sim6$.  They derive a lower limit 
on the optical depth of neutral hydrogen in the intergalactic medium,
from analysis of Ly$\alpha$ and Ly$\beta$, of 
$\tau(IGM) > 20$.

With Chandra, we detect 10 photons for SDS 1030+0524 
in the observed 0.3-6.5 $keV$ band, which is 2.2-47.0 keV in the quasar's
rest frame.  In fact all 10 photons have $E< 2.5 keV$ observed,
or $E<18$ keV.  The limits on IGM absorption are weak.  We
used Sherpa to fit the 10 photons 
with a model which had the Galactic absorption fixed,
a power law with fixed $\Gamma_x = 2.2$, 
and absorption at $z=6.28$. The column of
the redshifted absorption and normalization of the flux were free parameters.  
Significant
absorption was not detected, and we can place a 3$\sigma$ upper limit
to the absorbing column of 7.6$\times$10$^{23}$ atoms cm$^{-2}$ if the IGM 
has solar abundance, and 5.3$\times$10$^{24}$ atoms cm$^{-2}$ if
the IGM has hydrogen and helium only, at the solar ratio.  
In either case, the  
the Chandra data imply that at the Lyman limit, 
$\tau(IGM) < 10^6$ at redshift $\sim6$. 

\section{Discussion}

We have studied the evolution of quasars from $z=6$ to the present day,
putting together what is known about 
the optical continuum luminosities, broad emission line 
widths, and X-ray flux and spectrum. 
The data are consistent with a model where 
the optical-ultraviolet continuum arises in a cold, thin, optically
thick accretion disk, and the X-rays are produced at high redshift
by a Comptonized corona.  At low redshift, extra components are likely
making the X-ray spectrum more complex than the simple power law fits
available so far for most objects.

The high redshift quasars have systematically more massive central black holes
than their low-redshift counterparts, and are accreting at high rates, 
several tenths of the Eddington limit.  Their observed relatively weak X-ray
fluxes 
are a natural outcome of the accretion disk-coronal models, with no
cold absorption necessary to suppress the X-ray emission (c.f. Brandt et al. 
2001; Mathur 2001).

The Chandra observations require large values of the viscosity parameter, 
$\alpha > 0.02$ for the high redshift quasars.   
Numerical simulations show that the
turbulent part of the $\alpha$ parameter is negligible in comparison
to the magnetic term arising from magnetorotational instability and
$\alpha$ should be within 0.001-0.1 (Balbus \& Hawley 1998, 
Armitage 1998, Armitage et al 2001).  Observational constraints
based on AGN variability are limited (Siemiginowska \& Czerny 1989),
but in general they are in agreement with the theoretical predictions,
although for high disk luminosities $\alpha$ exceeds 0.1 
for the PG sample of quasars (Starling et al 2002).

We conclude that even at the highest redshifts probed to date, quasars 
were producing prodigious ultraviolet and soft-X-ray photons, which 
no doubt had interesting effects on the intergalactic medium and formation
of galaxies at the earliest times.   

\acknowledgements
We thank Harvey Tananbaum, Chandra director,
for using his director's time to observe
three of the high redshift quasars included in this study, and 
for putting the data in the public archive immediately.
We thank Harvey Tananbaum, Belinda Wilkes and Dan Schwartz for 
reading the first version of this paper carefully, and for 
suggestions which improved it.
This research is funded in part by NASA contracts NAS8-39073.
Support for this work was provided by the National Aeronautics and Space
Administration through Chandra Awards GO1-2117B, 
GO-01015A, and GO 12117A issued by the Chandra
X-Ray Observatory Center, which is operated by the Smithsonian Astrophysical
Observatory for and on behalf of NASA under contract NAS8-39073.
This research has made use of the NASA/IPAC 
Extragalactic Database (NED) which is operated by the Jet 
Propulsion Laboratory, California Institute of Technology, 
under contract with the National Aeronautics and Space Administration. 
Partial support was provided by NSF program AST-9617060.
Electronic files with the data used in Figures 1-3 are available at 
http://lithops.as.arizona.edu/$\sim$jill/ and http://hea-www.harvard.edu/QEDT/.

\begin{deluxetable}{lllcccccl}
%\scriptsize
%\rotate
%\tabletypesize{\scriptsize}
\tablecolumns{10}
\tablewidth{0pt}
\tablenum{1}
\tablecaption{Summary of Observations}
\tablehead{ 
\colhead{Quasar} & \colhead{$z_{\rm em}$} & \colhead{$AB ^a $} & \colhead{ObsID} & \colhead{Exp.} & \colhead{RA,Dec} & \colhead{$N_{\rm (Gal)}^b$} & \colhead{Net} & \colhead{Refn.} \\
\colhead{} & \colhead{} & \colhead{} & \colhead{Date} & 
\colhead{(sec)} & \colhead{J2000} & \colhead{} 
& \colhead{Counts$^c$} & \colhead{}} 
\startdata
PSS 0059+0003 & 4.178 & 19.45 & 2179 & 2682  & 00:59:22.80 & 3.20 & 12 & 1,2 \\
              &       &      & 18Sep2001 &   & +00:03:01.0 &      &     &     \\
BRI 0103+0032 & 4.437 & 18.84 & 2180  & 3709 & 01:06:19.20 & 3.10 & 26 & 1,2,3 \\
              &       &      &  18Sep2001 &  & +00:48:22.0 &      &      &     \\
SDS 0150+0041 & 3.67  & 18.35 & 2181  & 3238 & 01:50:48.80 & 2.79 & 7  & 8 \\
              &       &      &  31Aug2001 &  & +00:41:26.0 &      &      &     \\
BRI 0241-0146 & 4.053 & 18.45 & 875   & 7365 & 02:44:01.90 & 3.73 & 17 & 4   \\
              &       &      & 11Mar2000 &   & -01:34:03.0 &      &      &     \\
PSS 0248+1802 & 4.43  & 18.24 & 876   & 1731 & 02:48:54.30 & 9.72 & 19 & 1,2 \\
              &       &      & 27Dec1999 &   & +18:02:50.0 &      &      &     \\
BRI 0401-1711 & 4.236 & 18.84 & 2182  & 3841 & 04:03:56.60 & 2.34 & 15 & 4  \\
              &       &      & 3Aug2001 &   & -17:03:24.0 &      &      &  \\ 
SDS 0836+0054$^d$ & 5.82  & 18.8 & 3359 & 5687 & 08:36:43.85 & 4.15 & 24  & 9 \\
	      &       &      & 29Jan2002     &  & +00:54:53.3 &      &   &   \\
SDS 1030+0524 & 6.28  & 19.7 & 3357 & 7955  & 10:30:27.10 & 3.09 & 10  & 9 \\
              &       &      & 29Jan2002 &  & +05:24:55.1 &      &   &  \\
BRI 1033-0327 & 4.509 & 18.84 & 877 & 3447   & 10:36:23.70 & 4.85 & 16 & 5,6  \\
              &       &      & 26Jan2000 &   & -03:43:20.0 &      &      & \\
PSS 1057+4555 & 4.10  & 17.53 & 878  & 2808  & 10:57:56.40 & 1.17 & 34 & 2   \\
              &       &      & 14Jun2000 &   & +45:55:52.0 &      &      & \\
SDS 1204-0021 & 5.10  & 19.05 & 2183 & 1570  & 12:04:41.70 & 2.14 & 26 & 7   \\
              &       &      & 2Dec2000 &   & -00:21:49.0 &      &      &  \\
SDS 1306+0356 & 5.99  & 19.6 & 3358 & 8156 & 13:06:08.26 & 2.08 & 19  & 9 \\
              &       &      & 29Jan2002 &       & +03:56:26.3 &      &    &   \\
PSS 1317+3531 & 4.36  & 19.55 & 879 & 2788  & 13:17:43.20 & 0.99 & 9  & 2  \\
              &       &      & 14Jun2000 &  & +35:31:31.0 &      &      &     \\
PSS 1435+3057 & 4.35  & 19.12 & 880  & 2811 & 14:35:23.50 & 1.21 & $<9$  & 1,2 \\
              &       &      & 21May2000 &   & +30:57:23.0 &      &      &     \\ 
PSS 1443+2724 & 4.42  & 19.23 & 881  & 2170  & 14:43:31.20 & 2.33 & 10 & 1,2  \\
              &       &      & 12Jun2000 &   & +30:57:23.0 &      &      &     \\
SDS 1621-0042 & 3.70  & 17.41 & 2184 & 1570  & 16:21:16.90 & 7.29 & 26 & 7   \\
              &       &      & 5Sep2001 &   & -00:42:51.1 &      &     &     \\
BRI 2212-1626 & 3.99  & 18.65 & 2185 & 3222 & 22:15:27.20 & 2.64 & 15 & 4   \\
              &       &      & 16Dec2001 &  & -16:11:33.0 &      &      &     \\  
\enddata
%\footnotesize
\tablenotetext{a}{AB magnitude at 1450\AA ~ in quasar restframe, 
AB = -2.5 log $f_{opt}$ - 48.57, 
where [$f_{opt}$] = $erg~ cm^{-2}~ s^{-1}~ Hz^{-1}$. }
\tablenotetext{b}{Galactic H I column, in units 
of 10$^{20}$ atoms cm$^{-2}$, from {\sc COLDEN}.}
\tablenotetext{c}{Net number of photons in 
observed energy range 0.3 - 6.5 $keV$}
\tablenotetext{d}{SDS 0836+0054 is radio-loud -- see Fan et al. (2001b).}
\tablerefs{(1) Kennefick, J.\ D. et al. (1995); 
(2)Kennefick, J.\ D. et al.\ (1995);  
(3)Smith,J.\ D., et al.\ (1994); 
(4)Storrie-Lombardi,L.\ J., et al.\ (1996); 
(5)Storrie-Lombardi,L.\ J., et al.\ (1994); 
(6)Williger, G.\ M., et al.\ (1994);
(7)Fan, X., et al.\ (2000); 
(8)Fan, X., et al.\ (1999);
(9) Becker et al.\ (2001).}

\end{deluxetable}

\begin{deluxetable}{lccccccc}
%\tabletypesize{\scriptsize}
\scriptsize
%\begin{scriptsize}
%\tablecolumns{7}
%\tablewidth{0pt}
\tablenum{2}
\tablecaption{X-ray Results}
\tablehead{\colhead{QSO} & \colhead{$z_{\rm em}$} & \colhead{$\Gamma_x ^a$} 
& \colhead{$f_{\rm o}^b$} 
& \colhead{$log~f_{2keV}^c$} 
& \colhead{$log~f_{\rm 2500}^d$} 
& \colhead{$\alpha_{ox}^e$} 
& \colhead{$log~F_{\rm x}^f$}}
\startdata
PSS0059+0003 & 4.178 & 0.75$\pm$0.41& 3.38$\pm$1.25 & -31.150 & -27.137 & 1.540$^{+0.276}_{-0.236}$ & -13.17 \\
BRI0103+0032 & 4.437 & 1.78$\pm$0.30& 7.23$\pm$1.43 & -30.803 & -26.893 & 1.501$^{+0.245}_{-0.221}$& -13.58 \\
SDS0150+0041& 3.67  & 0.33$\pm$0.47&  1.17$\pm$0.44 & -31.668& -26.697 & 1.911$^{+0.259}_{-0.219}$& -13.26 \\
BRI0241-0146 & 4.053 & 1.33$\pm$0.36&  2.56$\pm$0.59 & -31.287& -26.737 & 1.749$^{+0.259}_{-0.219}$& -13.78 \\
PSS0248+1802 & 4.43  & 1.93$\pm$0.37& 14.78$\pm$3.48 & -30.488& -26.653 & 1.475$^{+0.252}_{-0.226}$& -13.38 \\
BRI0401-1711 & 4.236 & 1.28$\pm$0.37&  3.60$\pm$1.00 & -31.121& -26.893 & 1.625$^{+0.255}_{-0.226}$& -13.54 \\
SDS0836+0054 & 5.82  & 1.44$\pm$0.30& 4.30$\pm$0.93 & -30.906 & -26.877 & 1.549$^{+0.290}_{-0.260}$& -13.57\\  
SDS1030+0524 & 6.28  & 1.36$\pm$0.46& 1.21$\pm$0.58 & -31.422 & -27.237 & 1.609$^{+0.370}_{-0.305}$& -13.07\\ 
BRI1033-0327 & 4.509 & 2.39$\pm$0.42&  5.01$\pm$1.29 & -30.951& -26.964 & 1.560$^{+0.260}_{-0.232}$& -14.14 \\
PSS1057+4555 & 4.10 & 1.80$\pm$0.26& 11.47$\pm$1.97 & -30.631& -26.369 & 1.638$^{+0.228}_{-0.207}$& -13.39 \\
SDS1204-0021& 5.10 & 1.50$\pm$0.38&  2.30$\pm$0.61 & -31.236& -26.977 & 1.637$^{+0.280}_{-0.250}$& -13.89 \\
SDS1306+0356 & 5.99 & 1.32$\pm$0.33& 2.15$\pm$0.52 &  -31.194 & -27.197 & 1.537$^{+0.300}_{-0.269}$  & -13.79\\
PSS1317+3531 & 4.36 & 2.36$\pm$0.57&  2.83$\pm$1.01 & -31.213 & -27.177 & 1.552$^{+0.279}_{-0.240}$ & -14.36\\
PSS1435+3057 & 4.35 & \nodata     &  $<$3.0$^g$ & $<$-31.189   & -27.005 & $>$1.60 & \nodata\\  
PSS1443+2724 & 4.42 & 1.50$\pm$0.47&  4.47$\pm$1.46 & -31.009 & -27.049 & 1.522$^{+0.273}_{-0.238}$ & -13.60\\
SDS1621-0042& 3.70 & 1.39$\pm$0.30& 18.34$\pm$4.02 & -30.469 & -26.321 & 1.595$^{+0.222}_{-0.200}$ & -12.91\\
BRI2212-1626 & 3.99 & 1.27$\pm$0.37& 4.32$\pm$1.21  & -30.724 & -26.817 & 1.634$^{+0.247}_{-0.218}$ & -13.45 \\
\enddata
\tablenotetext{a}{$\Gamma_x$ = power law energy index, 
$A(E)=f_o[E/1 keV]^{-\Gamma_x}$.  Absorption fixed at Galactic NH}
\tablenotetext{b}{$f_{\rm o}$ = normalization of power law, in units of 
10$^{-6}$ $photons$ $keV^{-1}$ $s^{-1}$ $cm^{-2}$ at 1 $keV$. 
Note that f$_{\nu}$ (ergs cm$^{-2}$ sec$^{-1}$ Hz$^{-1}$)=6.63$\times10^{-27} f_o E(keV)$. } 
\tablenotetext{c}{f$_{2keV}$ = extrapolated flux at E=2 keV in the quasar rest frame,
erg cm$^{-2}$ s$^{-1}$ Hz$^{-1}$, assuming $\Gamma_x$=2.2.}
\tablenotetext{d}{$f_{2500}$ = 
extrapolated flux at 2500\AA, calculated from 
measured flux at 1450\AA, assuming power law with $\alpha$=-0.3; 
erg cm$^{-2}$ s$^{-1}$ Hz$^{-1}$.}
\tablenotetext{e}{$\alpha_{ox}$ = ratio of X-ray to optical flux, see equation (7) in 
text. Errors include uncertainties from tabulated uncertainties in $f_o$ and $\Gamma_x$ 
only.}
\tablenotetext{f}{Integrated observed 
flux within 2-10 keV, ergs cm$^{-2}$ s$^{-1}$.}
\tablenotetext{g}{Three sigma upper limit, assuming $\Gamma_x=2.2$.}
%\end{scriptsize}
\end{deluxetable}

\clearpage
\begin{deluxetable}{llccc}
%\scriptsize
%\rotate
%\tabletypesize{\scriptsize}
\tablecolumns{10}
\tablewidth{0pt}
\tablenum{3}
\tablecaption{X-ray Luminosities}
\tablehead{ 
\colhead{Quasar} & \colhead{$z_{\rm em}$} & \colhead{$log L_x$$^a$} 
& \colhead{$log L_x$$^a$} 
& \colhead{$log L_x$$^a$}\\ 
\colhead{} & \colhead{} & \colhead{$q_0=0$$^b$} & \colhead{$q_0=1/2$$^c$} & 
\colhead{$\Lambda CDM$$^d$} 
} 
\startdata
PSS 0059+0003 & 4.178 & 46.39  & 45.70  & 46.06  \\
BRI 0103+0032 & 4.437 & 46.07  & 45.35  & 45.71  \\
SDS 0150+0041 & 3.67  & 46.11  & 45.48  & 45.83  \\
BRI 0241-0146 & 4.053 & 45.74  & 45.06  & 45.42   \\
PSS 0248+1802 & 4.43  & 46.27  & 45.54  & 45.91   \\
BRI 0401-1711 & 4.236 & 46.04  & 45.34  & 45.70   \\
SDS 0836+0054 & 5.82  & 46.48  & 45.62  & 46.00  \\
SDS 1030+0524 & 6.28  & 47.10  & 46.19  & 46.58  \\
BRI 1033-0327 & 4.509 & 45.53  & 44.80  & 45.17  \\
PSS 1057+4555 & 4.10  & 46.14  & 45.46  & 45.82   \\
SDS 1204-0021 & 5.10  & 45.97  & 45.17  & 45.54   \\
SDS 1306+0356 & 5.99  & 46.30  & 45.43  & 45.81   \\
PSS 1317+3531 & 4.36  & 45.26  & 44.55  & 44.91   \\
PSS 1443+2724 & 4.42  & 46.04  & 45.32  & 45.69   \\
SDS 1621-0042 & 3.70  & 46.48  & 45.84  & 46.20   \\
BRI 2212-1626 & 3.99  & 46.04  & 45.37  & 45.73   \\
\enddata
%\footnotesize
\tablenotetext{a}{X-ray luminosity, in observed 2-10 keV band, in ergs s$^{-1}$}
\tablenotetext{b}{q$_0$=0, H$_0$=70 km s$^{-1}$ Mpc$^{-1}$, $\Lambda=0$.}
\tablenotetext{c}{q$_0$=1/2, H$_0$=70 km s$^{-1}$ Mpc$^{-1}$, $\Lambda=0$}
\tablenotetext{d}{$\Omega_{\Lambda} = 0.7$, $\Omega_{M}=0.3$, 
H$_0$=70 km s$^{-1}$ Mpc$^{-1}$}
\end{deluxetable}

\clearpage
\begin{figure}
%\epsscale{0.5}
\includegraphics[height=6.8in,width=6.in]{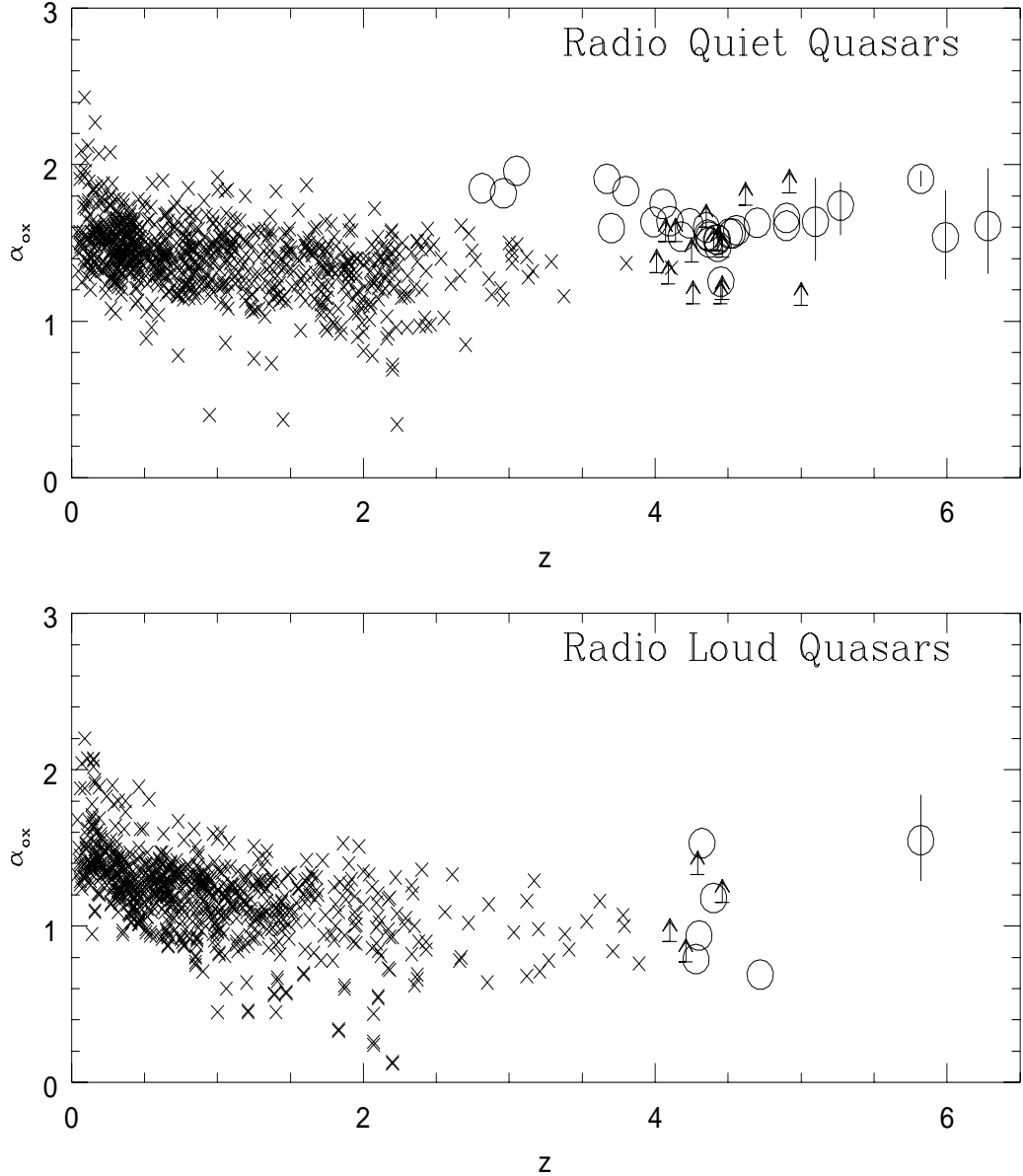}
%\plotone{f1.eps}
\caption{X-ray to optical flux ratio versus redshift, $z$, 
for radio-quiet (top) and 
radio-loud (bottom) quasars.  X's are quasars detected by ROSAT, from 
Yuan et al. (1998) 
for radio quiet objects and Brinkman et al. (1997) for radio loud quasars;  
Open circles are high redshift quasars observed with Chandra (see text).}
\end{figure}

\clearpage
\begin{figure}
%\epsscale{0.85}
\includegraphics[height=6.8in,width=6.in]{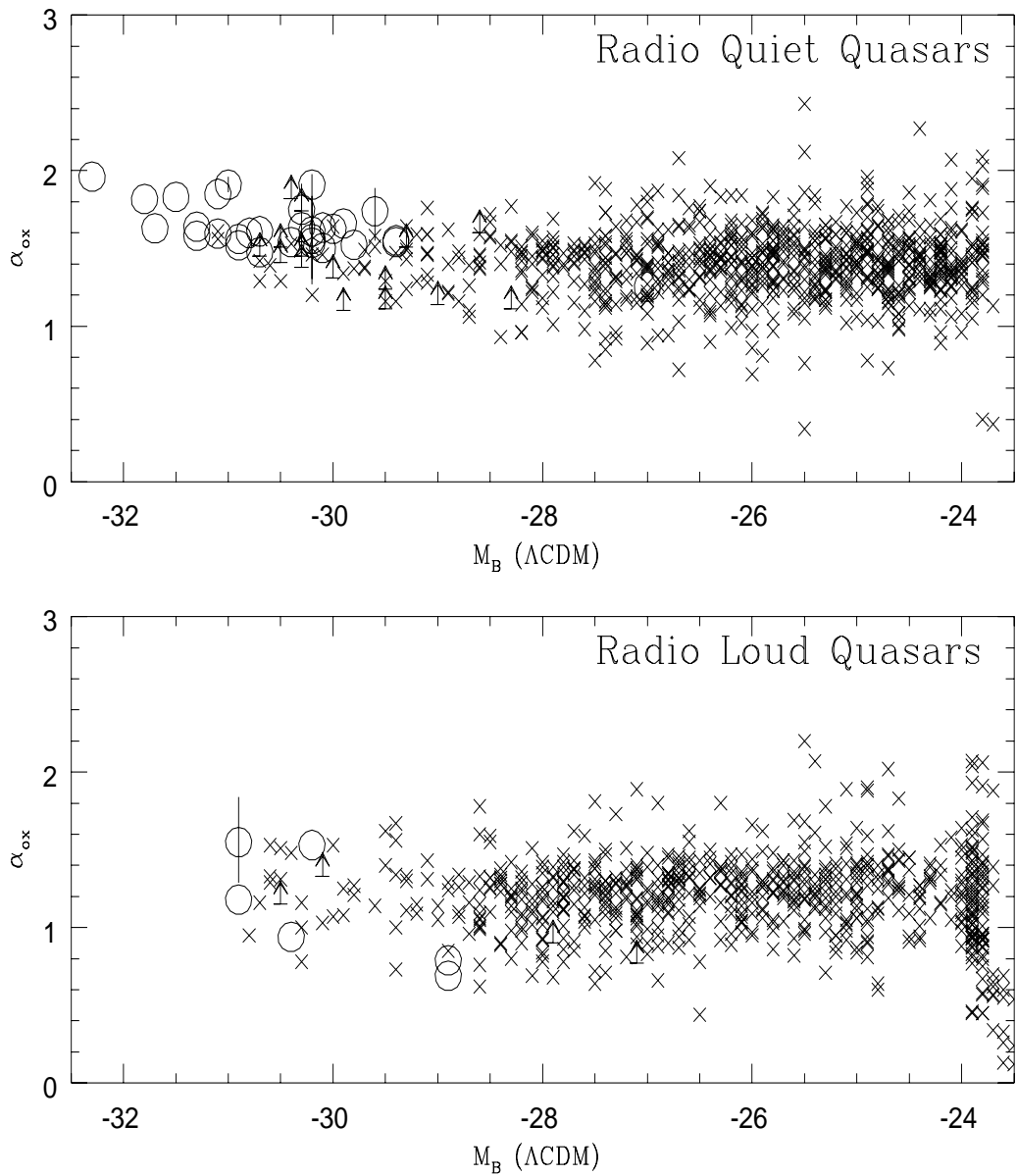}
%\plotone{f2.eps}
\caption{X-ray to optical flux ratio versus absolute B-magnitude, M$_{B}$, 
computed assuming $\Lambda$CDM cosmology.  
X's are radio
quiet quasars from Yuan et al. (1998) and radio-loud quasars from
Brinkman et al. (1997).  Open circles are high
redshift quasars observed with Chandra (see text).}
\end{figure}

\clearpage
\begin{figure}
\includegraphics[height=6.8in,width=6.in]{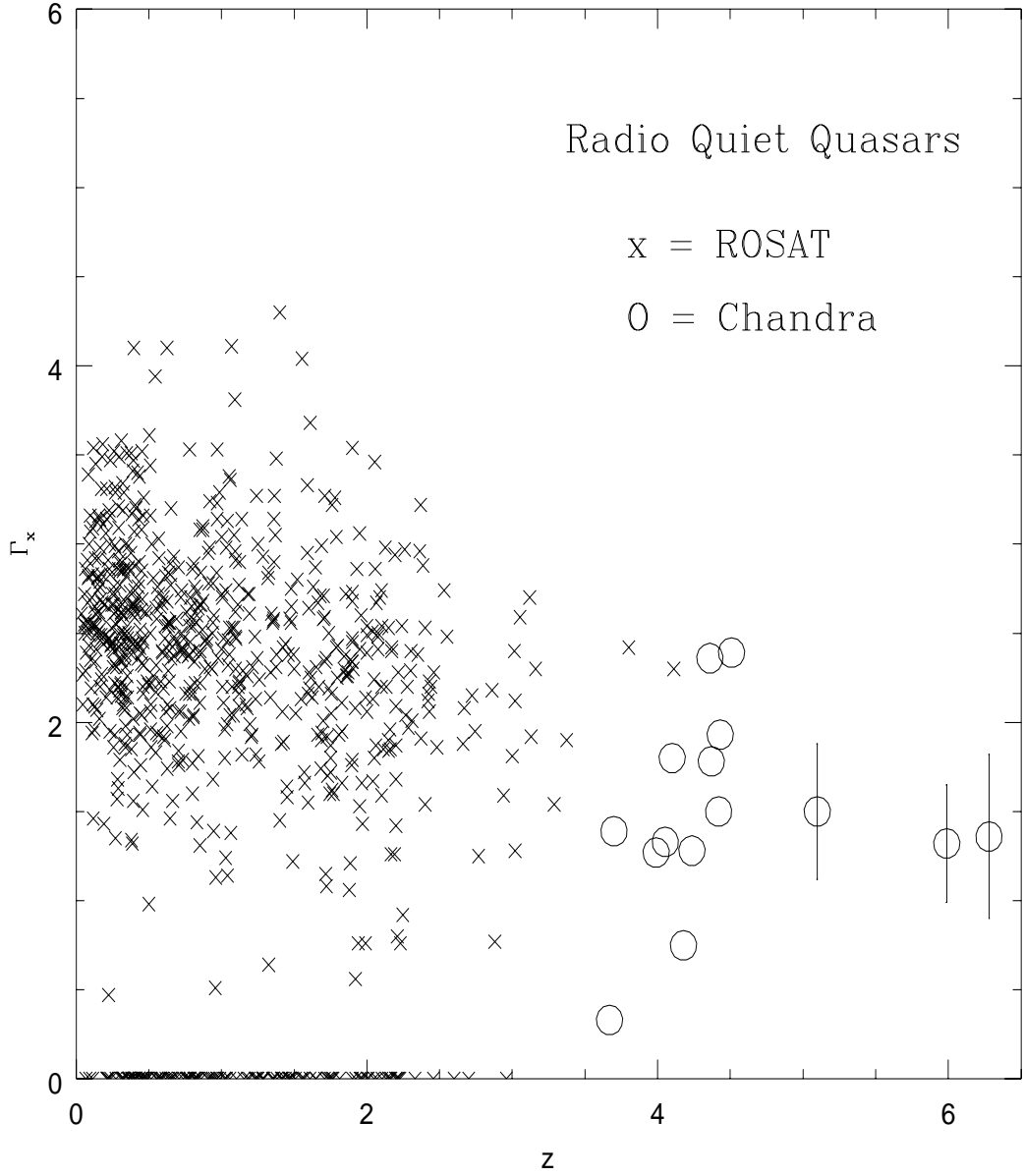}
%\plotone{f3.eps}
\caption{X-ray spectral index, $\Gamma_x$, versus versus redshift, $z$.  
X's are radio
quiet quasars from Yuan et al. (1998);  Open circles are high
redshift quasars (see text). For quasars with too few counts detected
to derive $\Gamma_x$ we plot $\Gamma_x$=0.}
\end{figure}

\clearpage
\begin{figure}
\includegraphics[height=6.8in,width=6.in]{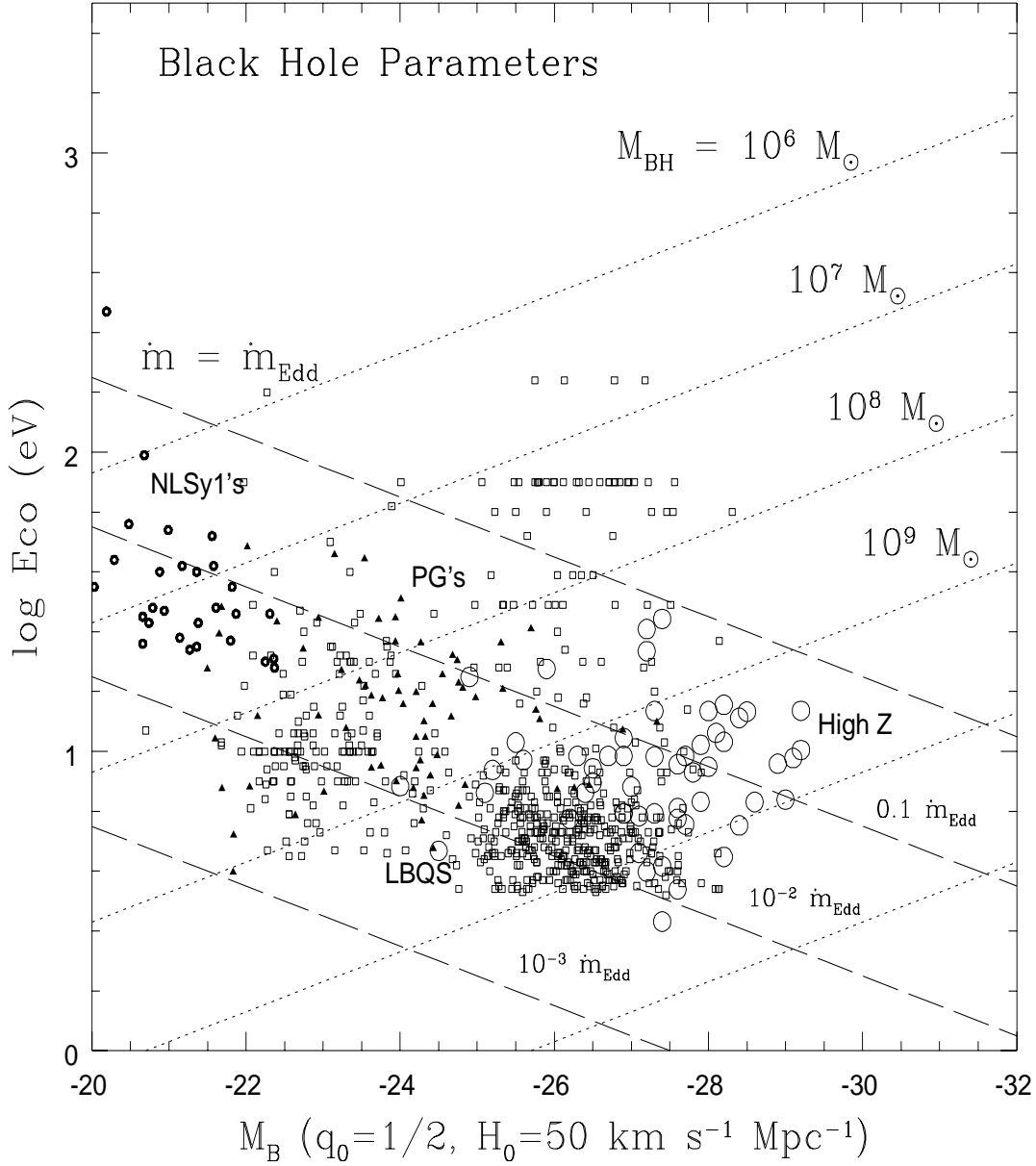}
%\plotone{f4.eps}
\caption{ Black hole parameters.  Plotted is the UV cut-off energy, E$_{CO}$ in eV, 
versus absolute B magnitude, M$_{B}$, for $q_0 = 0.5$, $\Lambda$=0, and 
H$_{0}$ = 50 km s$^{-1}$ Mpc$^{-1}$.  Small open 
circles are Narrow Line Sy1's, triangles 
are PG quasars, open squares 
are LBQS quasars, and large open circles are high redshift quasars.
Dotted lines show locus of constant black hole mass, for 
$M_{BH}=10^6,10^7,10^8,10^9,10^{10}$, and $10^{11}$ M$_{\odot}$.  
Long dashed lines are locus of constant mass accretion
rate, in units of 1.0, 0.1, 0.01 and 0.001 the Eddington rate.
}
\end{figure}

\clearpage
\begin{figure}
\includegraphics[height=6.8in,width=6.in]{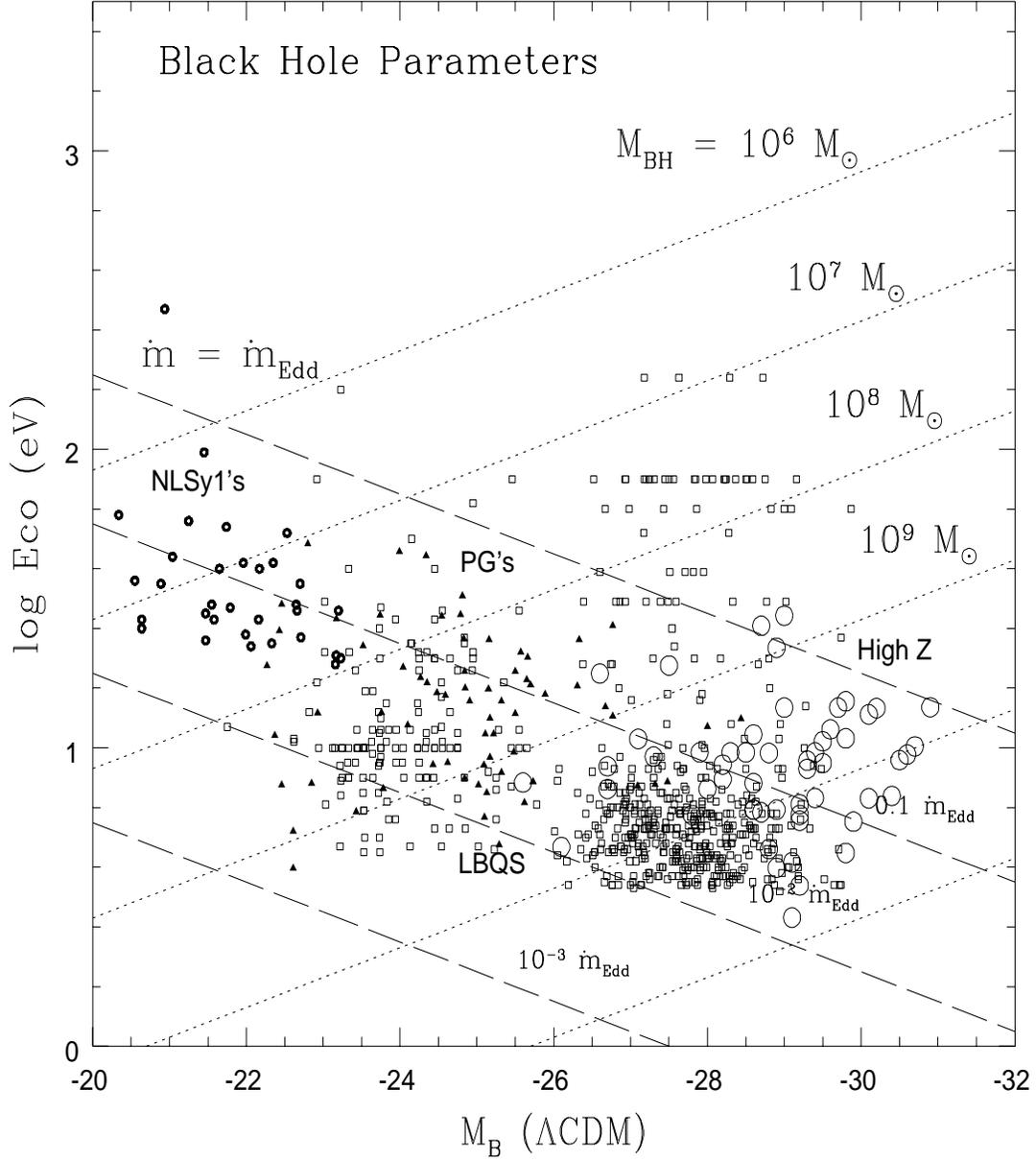}
\caption{
Same as figure 4, except $M_B$ was computed assuming $\Lambda$CDM cosmology.
}
\end{figure}

\clearpage
\begin{figure}
\includegraphics[height=6.8in,width=6.in]{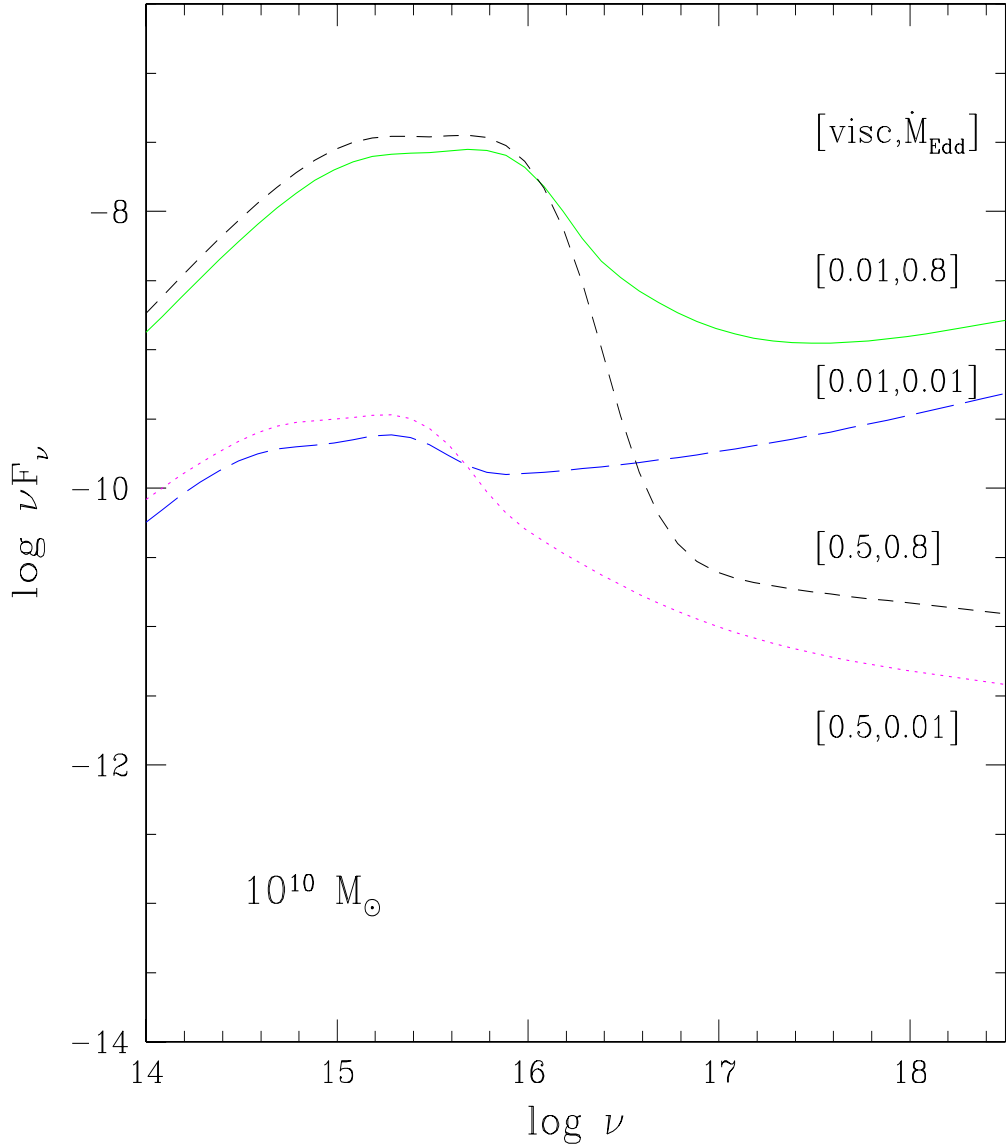}
\caption{
Representative spectral energy distributions for two-phase accretion disk models of
Janiuk \& Czerny (2000).  We assume a $10^{10}$ M$_{\odot}$ black hole, and viscosity 
parameter $\alpha$=0.01 and 0.5, for mass accretion rate = 0.8 and 0.01 times 
the Eddington limit.  
}
\end{figure}
\clearpage
\begin{figure}
\includegraphics[height=6.8in,width=6.in]{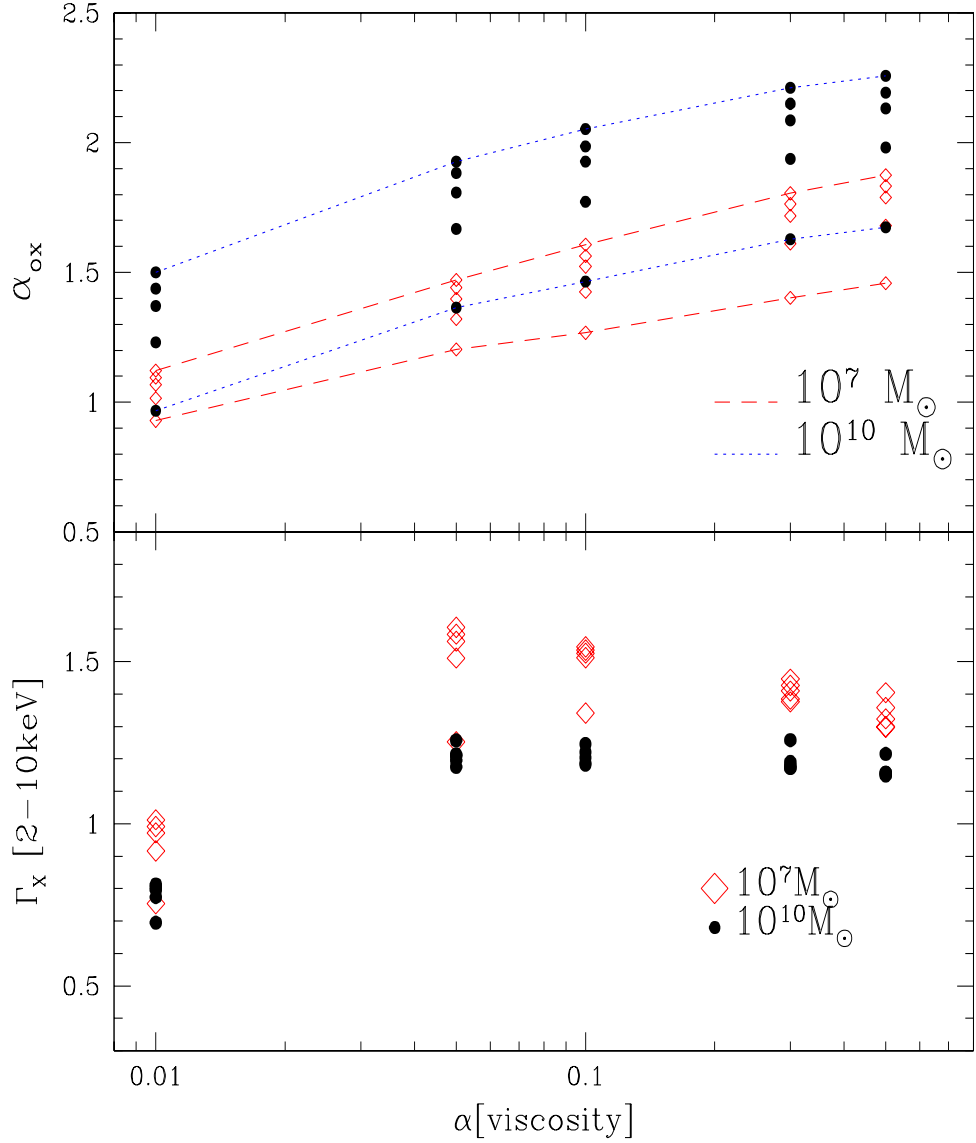}
\caption{
Predicted dependence of $\alpha_{ox}$ (top panel) and 
X-ray power law spectral index $\Gamma_x$ (bottom panel) on viscosity, $\alpha$, for two-phase accretion 
disk models of Janiuk \& Czerny (2000).
Dotted lines are for a supermassive black hole with $10^{10}$ M$_{\odot}$,
dashed lines have $10^{7}$ M$_{\odot}$ black hole.  Points are plotted for
accretion rates of 0.01, 0.1, 0.3, 0.5, 0.8 times the Eddington limit.
}
\end{figure}

\end{document}